\documentclass{webofc}
\usepackage[varg]{txfonts}   % Web of Conferences font
\usepackage{hyperref}
\usepackage{url}

\usepackage{bm}
\usepackage{float}
\usepackage{placeins}
\usepackage[captionskip=-2pt]{subfig}
\usepackage{upgreek}
\usepackage[hang,flushmargin]{footmisc}
\usepackage{parskip}
\usepackage{lineno}
\hypersetup{colorlinks=true,citecolor=blue,urlcolor=blue,linkcolor=blue}
\begin{document}

\title{Top-Quark Pair Production in Heavy-Ion \\ Collisions in the ATLAS Experiment}
\shorttitle{Top-Quark Pair Production in Heavy-Ion Collisions}

\raggedbottom

\author{\firstname{Patrycja} \lastname{Potępa}\inst{1,2}\fnsep\thanks{Speaker} on behalf of the ATLAS Collaboration
}
\shortauthor{P. Potępa}

\institute{
AGH University of Krakow
\and
Johannes Gutenberg University Mainz
}

\abstract{%
  Top-quark pair production in heavy-ion collisions provides a unique opportunity to probe nuclear parton distribution functions and study the time evolution of strongly interacting matter, including the quark-gluon plasma. This work presents the observation and measurement of top-quark pair production in both proton--lead ($p$+Pb) and lead--lead (Pb+Pb) collisions using the ATLAS experiment at the Large Hadron Collider (LHC). In $p$+Pb collisions at a centre-of-mass energy of 8.16 TeV, top-quark pair production is observed in the lepton+jets and dilepton channels, with significances exceeding 5 standard deviations in each channel. The nuclear modification factor, $R_{p\mathrm{A}}$, is measured for the first time in this process, providing new insights into nuclear parton distribution functions. In Pb+Pb collisions at a centre-of-mass energy of 5.02~TeV, top-quark pair production is studied in the ($e\mu$) final state, using datasets recorded in 2015 and 2018 with an integrated luminosity of 1.9~nb$^{-1}$. The measurement achieves a significance of 5.0 standard deviations and is compared to theoretical predictions based on various nuclear PDF sets. These measurements establish top-quark pairs as valuable tools for investigating heavy-ion collisions, offering novel insights into the dynamics of the quark-gluon plasma and nuclear matter.
}

\Conference{Presented at the 32nd International Symposium on Lepton Photon Interactions at High Energies, Madison, Wisconsin, USA, August 25-29, 2025}
\maketitle

\begingroup
\renewcommand\thefootnote{}\footnote{
Copyright 2025 CERN for the benefit of the ATLAS Collaboration. Reproduction of this article or parts of it is allowed as specified in the CC-BY-4.0 license.
}%
\addtocounter{footnote}{-1}%
\endgroup

\newpage
%-------------------------------------------------------------------------------
\section{Top-quark pair production in \textit{p}+Pb collisions}
\label{sec:pPb}
%-------------------------------------------------------------------------------

Top-quark production in proton--lead~($p$+Pb) collisions offers a new means to probe nuclear parton distribution functions~(nPDFs) in a poorly constrained kinematic regime~\cite{bib:nPDFttbar}. Differences between $p$+Pb and proton--proton ($pp$) systems can be quantified through the nuclear modification factor, $R_{p\mathrm{A}}$. Data from $p$+Pb collisions collected during Run~2 with the ATLAS detector~\cite{bib:atlas} provide experimental access to top-quark production yields. The data were collected in 2016 at a nucleon--nucleon centre-of-mass energy of $\sqrt{s_{\mathrm{NN}}}=8.16$~TeV, and correspond to an integrated luminosity of 165~nb$^{-1}$.

\begin{figure}[!b]   
\centering
\subfloat[]{\includegraphics[width=0.3\textwidth]{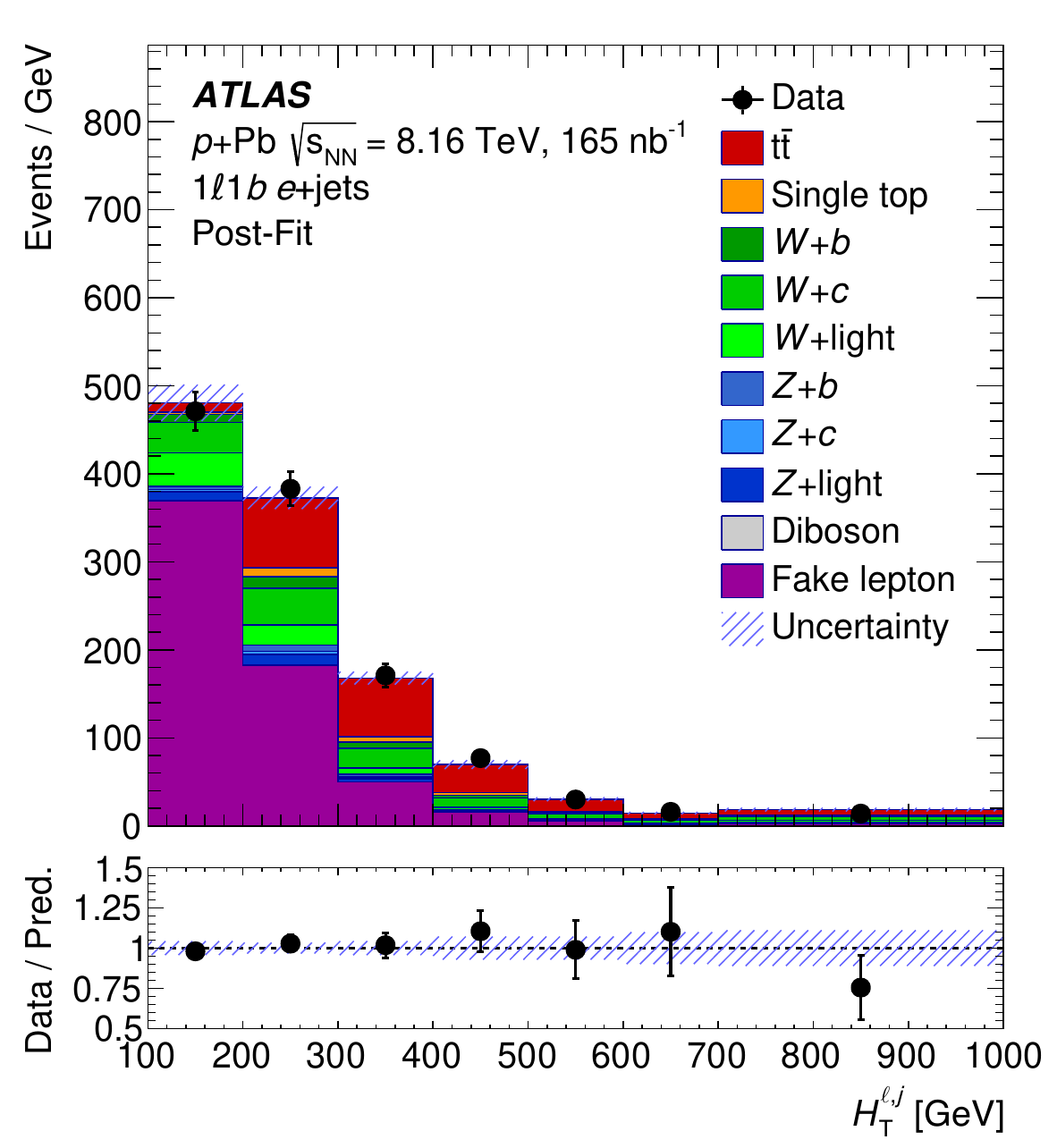}}
\subfloat[]{\includegraphics[width=0.3\textwidth]{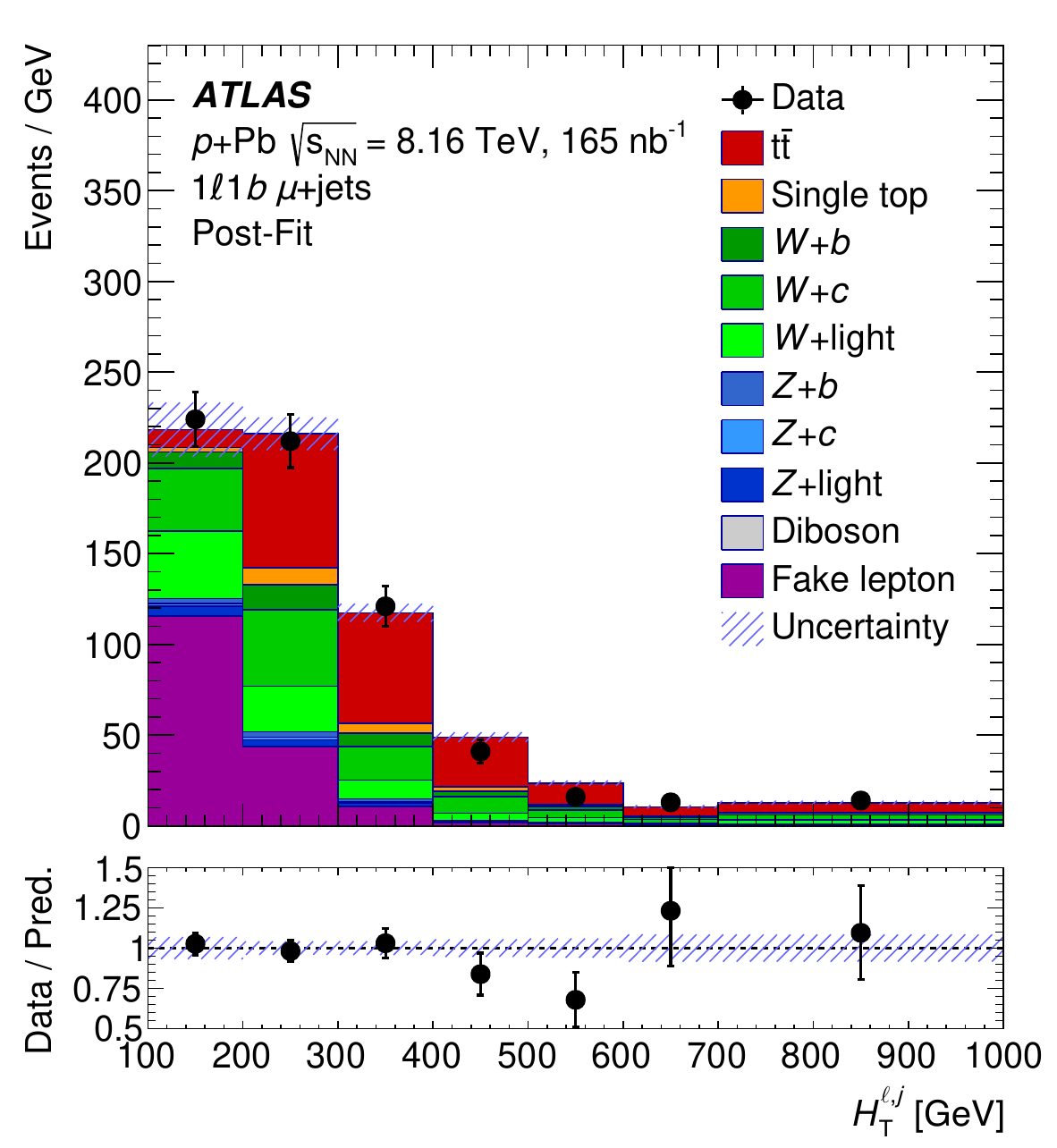}}
\subfloat[]{\includegraphics[width=0.3\textwidth]{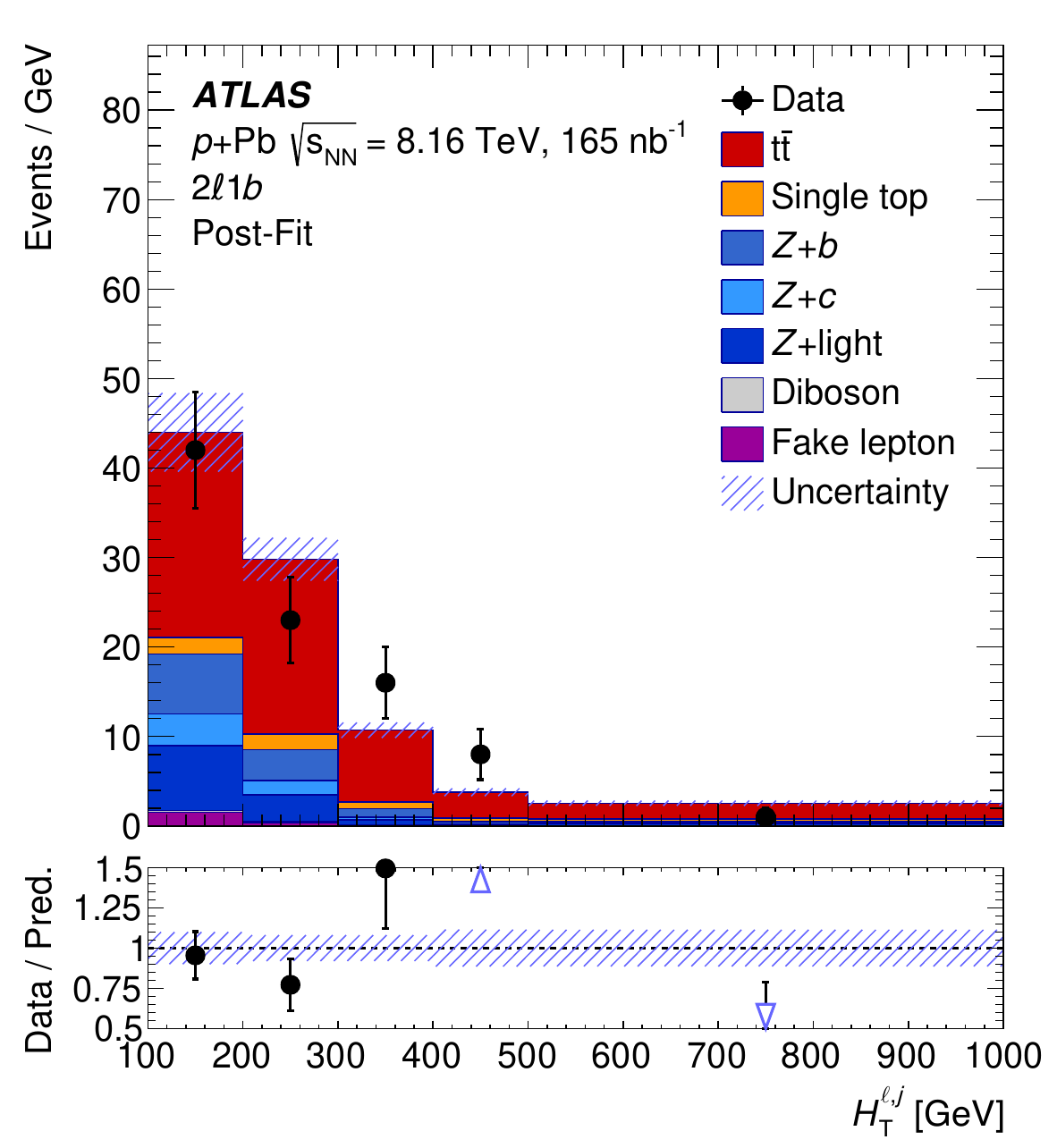}} \\
\subfloat[]{\includegraphics[width=0.3\textwidth]{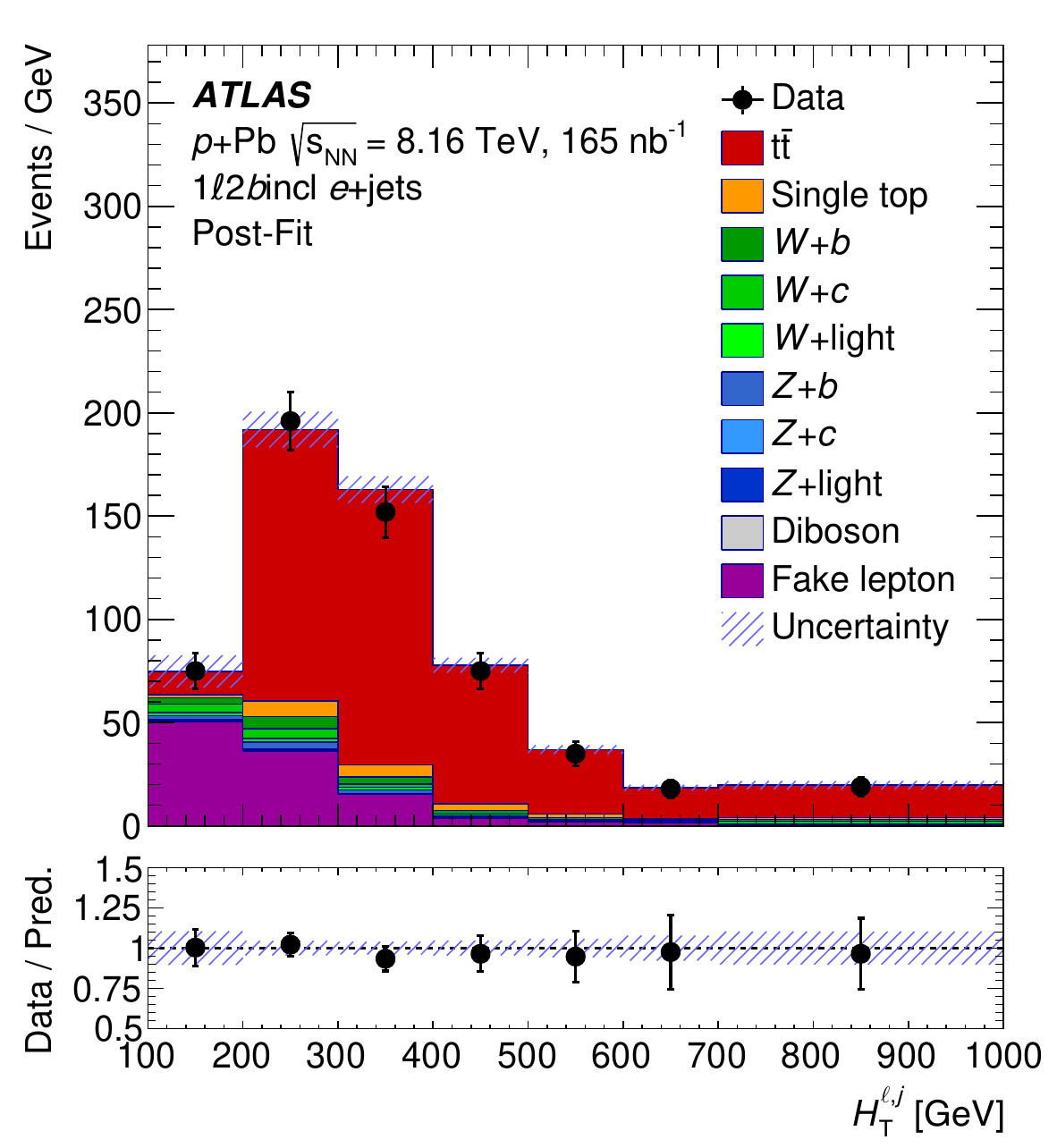}}
\subfloat[]{\includegraphics[width=0.3\textwidth]{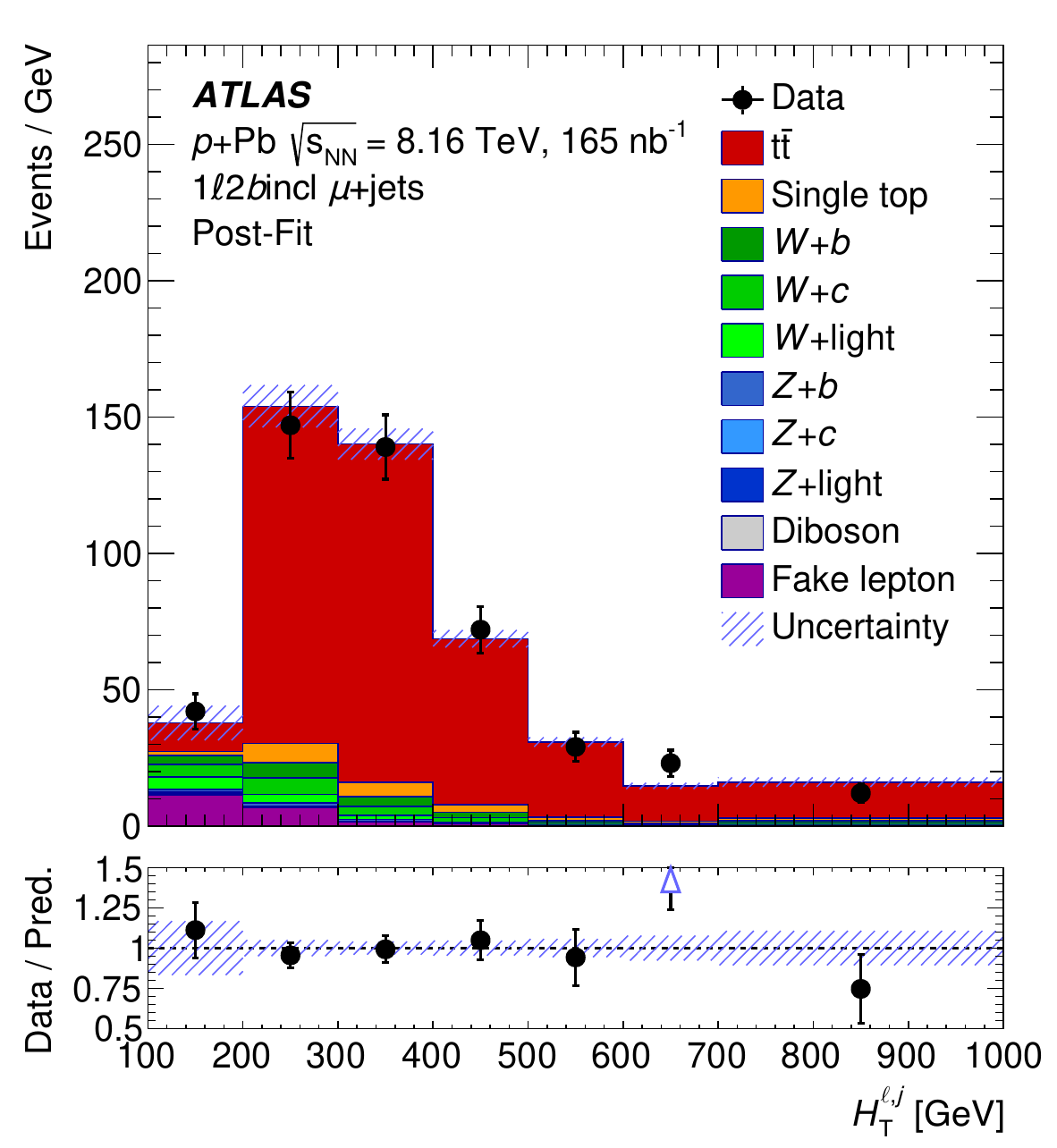}}
\subfloat[]{\includegraphics[width=0.3\textwidth]{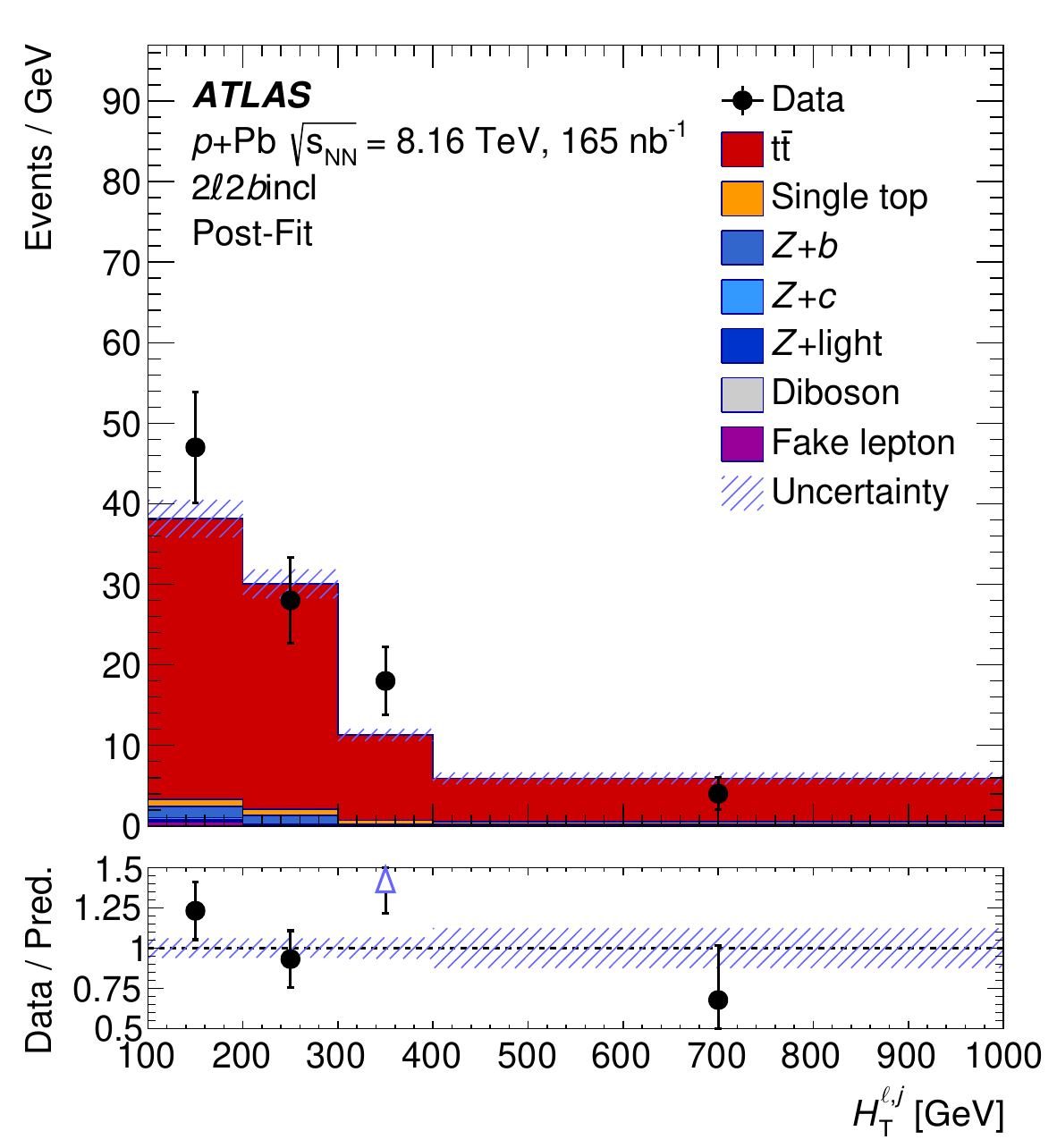}}
\caption{Post-fit plots representing the $H_\mathrm{T}^{\ell,j}$ variable in each of the six signal regions ($e$+jets: (a)~$1\ell1b$ and (d)~$1\ell2b\mathrm{incl}$, $\mu$+jets: (b)~$1\ell1b$ and (e)~$1\ell2b\mathrm{incl}$, dilepton: (c)~$2\ell1b$ and (f)~$2\ell2b\mathrm{incl}$), with total uncertainties represented by the hatched area. The bottom panels show ratios of data and a sum of predictions~\cite{bib:ttbarpPb}.}
\label{fig:postfitpPb}
\end{figure}

Events containing electrons or muons are used to reconstruct top-quark pairs. The selection relies on single-electron and single-muon triggers with a minimum transverse momentum~($p_\mathrm{T}$) threshold of 15~GeV~\cite{bib:trigger}. Electrons (muons) must have $p_\mathrm{T}>18$~GeV and $|\eta|<2.47~(2.5)$, and meet ‘Medium’ identification and isolation criteria~\cite{bib:electron,bib:muon}. Jets are built using the anti-$k_t$ algorithm~\cite{bib:antikt} with a radius of $R=0.4$ and must have $p_\mathrm{T}>20$~GeV and $|\eta|<2.5$, while $b$-jets are identified using the DL1r algorithm~\cite{bib:DL1r}. The fake-lepton background is estimated from data using the Matrix Method~\cite{bib:fakes}. Six signal regions are constructed using the scalar sum of the $p_\mathrm{T}$ of leptons and jets~($H_\mathrm{T}^{\ell j}$) as illustrated in Figure~\ref{fig:postfitpPb}. The $\ell$+jets channel is divided into four regions, corresponding to events with one electron or muon and either exactly one or at least two $b$-tagged jets, denoted as $1\ell1b~e$+$\mathrm{jets}$, $1\ell2b\mathrm{incl}~e$+$\mathrm{jets}$, $1\ell1b~\mu$+$\mathrm{jets}$, and $1\ell2b\mathrm{incl}~\mu$+$\mathrm{jets}$. Two additional signal regions are defined in the dilepton channel, requiring exactly one or at least two $b$-jets, labelled $2\ell1b$ and $2\ell2b\mathrm{incl}$, respectively.

A profile-likelihood fit method~\cite{bib:fit} is used to extract the signal strength~($\mu_{t\bar{t}}$), defined as the ratio of the observed $t\bar{t}$ cross section to the Standard Model expectation. The $\mu_{t\bar{t}}$ value is converted into the inclusive $t\bar{t}$ production cross section in $p$+Pb collisions at $\sqrt{s_\mathrm{NN}}=8.16$~TeV, which amounts to $\sigma_{t\bar{t}}^{p+\mathrm{Pb}} = 58.1\pm 2.0\;\mathrm{(stat.)}\;^{+4.8}_{-4.4} \;\mathrm{(syst.)}\;\mathrm{nb} = 58.1\;^{+5.2}_{-4.9} \;\mathrm{(tot.)}\;\mathrm{nb}$. The total relative uncertainty of 9\% results in the most precise $t\bar{t}$ cross-section measurement in heavy-ion collisions to date.

The nuclear modification factor~($R_{p\mathrm{A}}$) is used to quantify differences between $p$+Pb and $pp$ collisions. It is defined as the ratio of the $t\bar{t}$ cross section in $p$+Pb collisions to that in $pp$ collisions, scaled by the lead mass number ($A_\mathrm{Pb}$=208). $R_{p\mathrm{A}}$ for the $t\bar{t}$ process is measured to be $R_{p\mathrm{A}} = 1.090\pm0.039\;\mathrm{(stat.)}\;^{+0.094}_{-0.087}\;\mathrm{(syst.)} = 1.090\pm0.100\;\mathrm{(tot.)}$. The result is consistent with the expectation from scaled $pp$ collisions within one standard deviation.

\begin{figure}[H]
\centering
\subfloat[]{\includegraphics[width=0.47\textwidth]{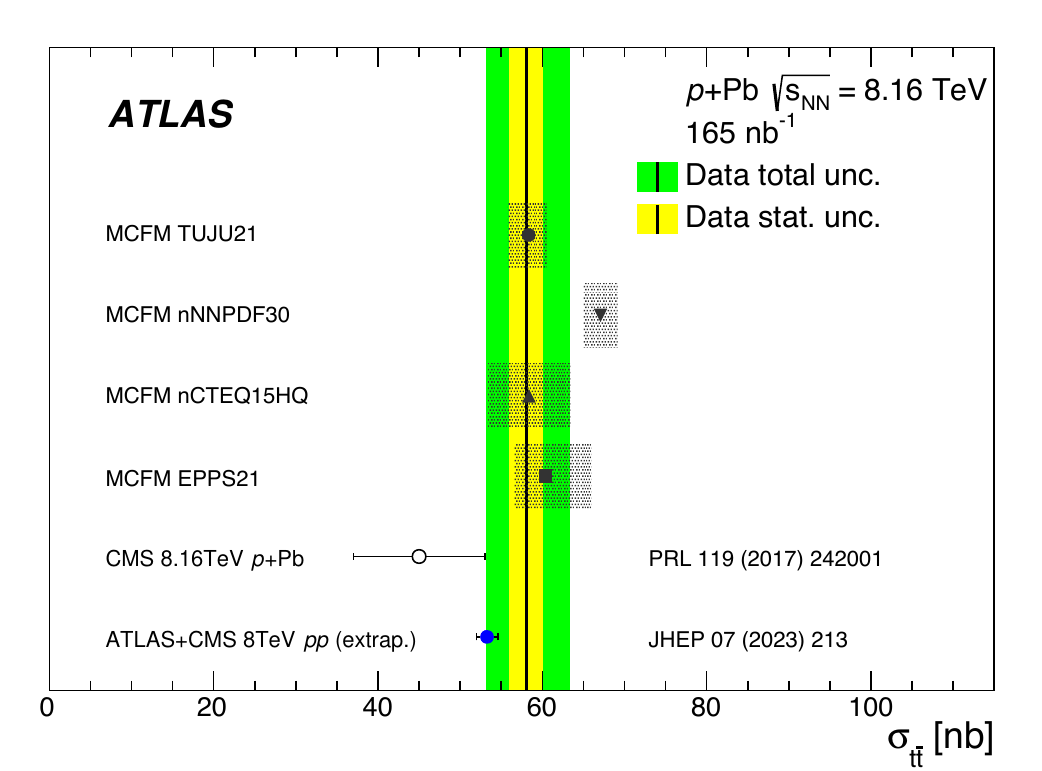}} \hfill
\subfloat[]{\includegraphics[width=0.47\textwidth]{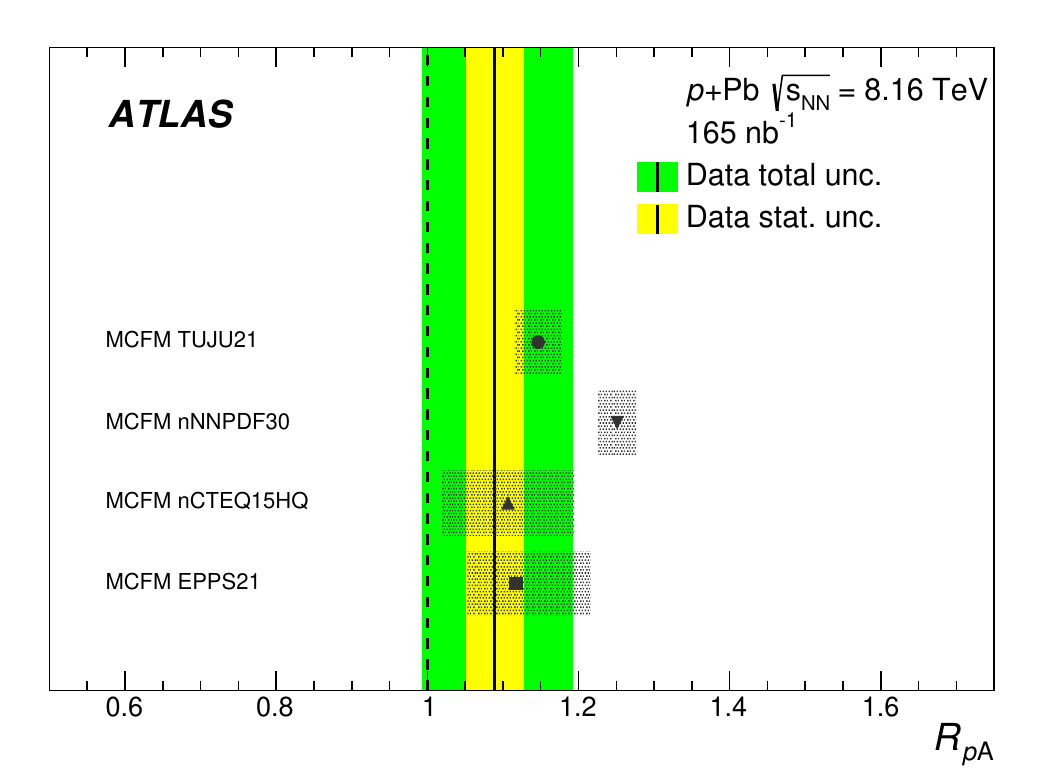}}
\caption{Comparison of the (a)~cross section and (b)~nuclear modification factor for $t\bar{t}$ production with theoretical models and other measurements. The dashed line at $R_{p\mathrm{A}}=1$ indicates the geometric expectation based on $pp$ collisions~\cite{bib:ttbarpPb}.}
\label{fig:XSpPb}
\end{figure}

The measured $\sigma_{t\bar{t}}^{p+\mathrm{Pb}}$ and $R_{p\mathrm{A}}$ are compared with previous experimental results and theoretical predictions in Figure~\ref{fig:XSpPb}. The cross section is in agreement with the CMS measurement in $p$+Pb collisions~\cite{bib:ttbarpPbCMS}, and the combined $pp$ cross section at $\sqrt{s}=8$~TeV by ATLAS and CMS~\cite{bib:ttbarpp8TeV}, scaled by $A_\mathrm{Pb}$ and extrapolated to the corresponding energy. Both $\sigma_{t\bar{t}}^{p+\mathrm{Pb}}$ and $R_{p\mathrm{A}}$ are consistent with predictions using four nPDF sets: TUJU21~\cite{bib:TUJU21}, nNNPDF3.0~\cite{bib:nNNPDF30}, nCTEQ15HQ~\cite{bib:nCTEQ15HQ}, and EPPS21~\cite{bib:EPPS21}.

%-------------------------------------------------------------------------------
\section{Top-quark pair production in Pb+Pb collisions}
\label{sec:PbPb}
%-------------------------------------------------------------------------------

As the heaviest particle carrying colour charge, the top quark offers a unique opportunity to probe the properties of the quark-gluon plasma (QGP)~\cite{bib:QGP}, in particular its time evolution~\cite{bib:QGPtime}. The first observation of top-quark pair production in Pb+Pb collisions~\cite{bib:ttbarPbPb} was enabled by the full Pb+Pb dataset collected by ATLAS in Run~2. The data were recorded at a nucleon--nucleon centre-of-mass energy of $\sqrt{s_{\mathrm{NN}}}=5.02$~TeV, and yield an integrated luminosity of 1.9~nb$^{-1}$.

\begin{figure}[H]   
\centering
\subfloat[]{\includegraphics[width=0.35\textwidth]{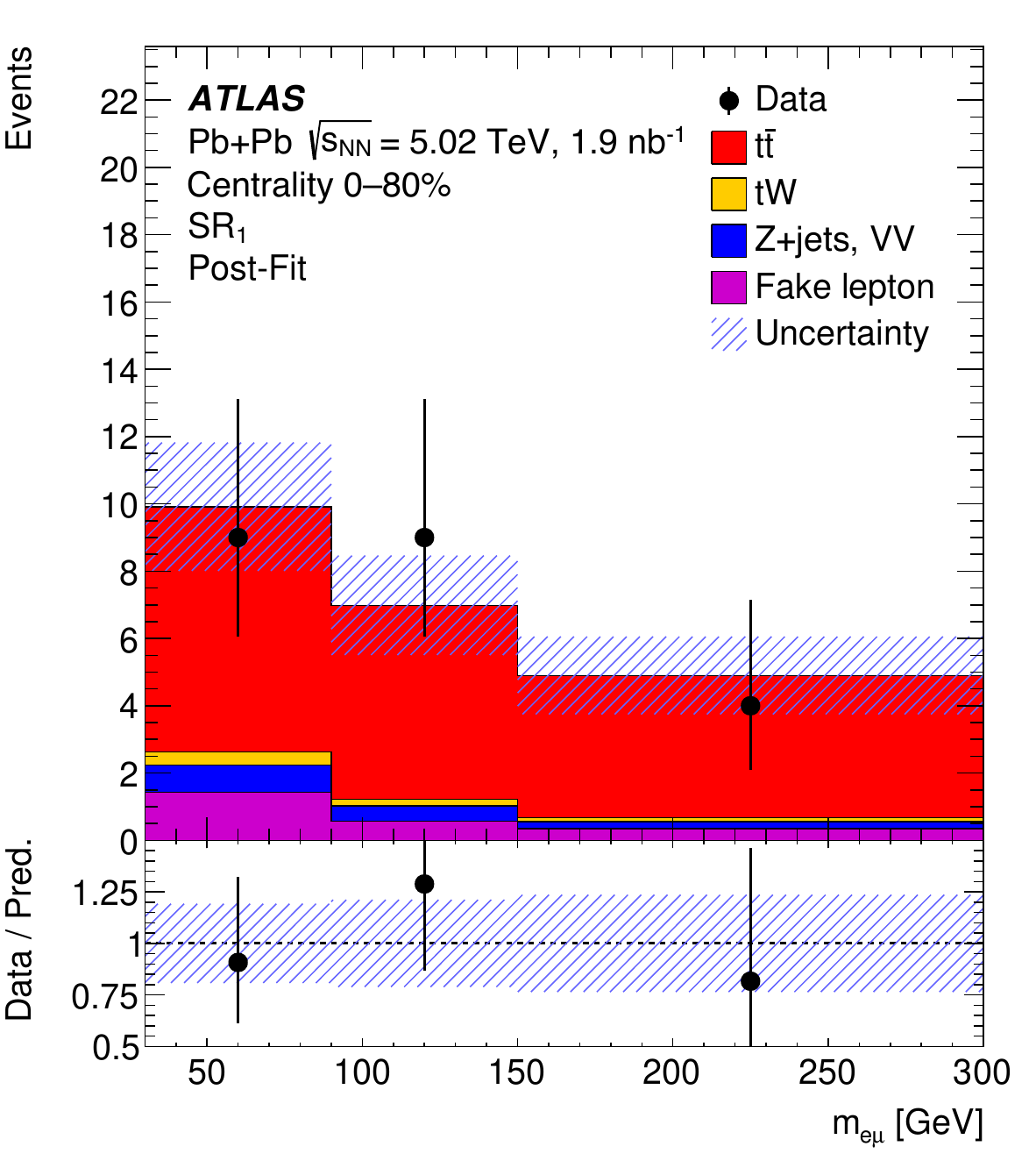}} \hspace{20pt}
\subfloat[]{\includegraphics[width=0.35\textwidth]{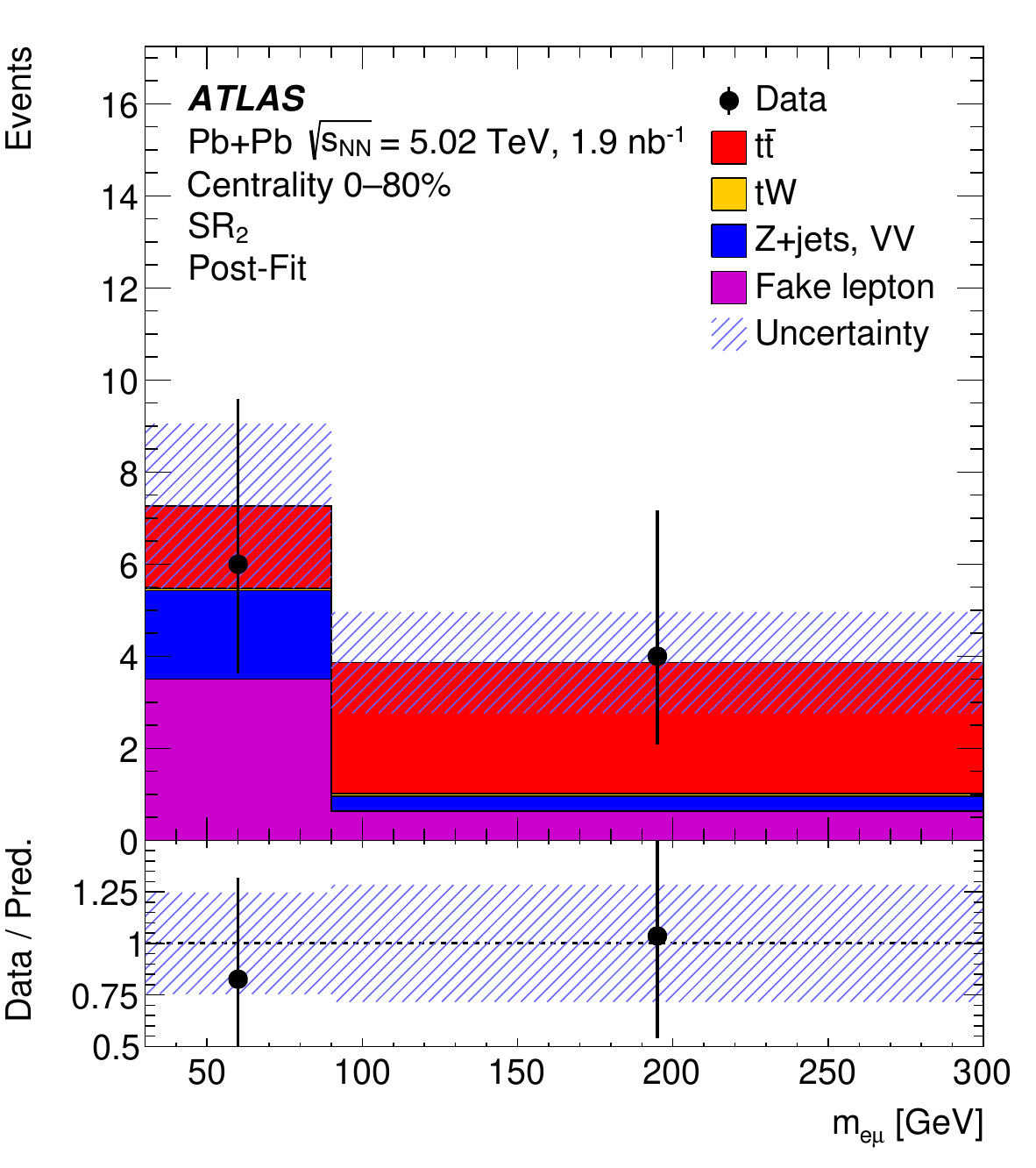}}
\caption{Post-fit plots representing the $m_{e\mu}$ variable in (a)~SR$_1$ and (b)~SR$_2$, with total uncertainties represented by the hatched area. The bottom panels show ratios of data and a sum of predictions~\cite{bib:ttbarPbPb}.}
\label{fig:postfitPbPb}
\end{figure}

The measurement focuses on events in the electron--muon ($e\mu$) channel of $t\bar{t}$ decays in the 0--80\% centrality interval, with exactly one oppositely charged $e\mu$ pair. Single-electron and single-muon triggers are employed, with $p_\mathrm{T}$ thresholds of 15 and 8~GeV, respectively~\cite{bib:trigger}. Electron (muon) candidates are required to have $p_\mathrm{T}>18$ (15) GeV and $|\eta|<2.47$ (2.5), and must satisfy the `Loose’ identification and isolation criteria~\cite{bib:electron,bib:muon}. Jets are reconstructed from massless calorimeter towers with the anti-$k_t$ algorithm~\cite{bib:antikt} using a radius of $R=0.4$, and must fulfil $p_\mathrm{T}>35$~GeV and $|\eta|<2.5$. The contribution from the underlying-event background is subtracted on an event-by-event basis. The fake-lepton contribution is estimated using data-driven methods. Two signal regions (SR$_1$ and SR$_2$) are defined: events with $p_\mathrm{T}^{e\mu}>40$~GeV are assigned to SR$_1$, while those with $p_\mathrm{T}^{e\mu}\leq40$~GeV form SR$_2$. Figure~\ref{fig:postfitPbPb} shows the comparison between the observed $m_{e\mu}$ distributions and the combined signal and background predictions in the two regions.

\begin{figure}[!b]
\centering
\includegraphics[width=0.515\textwidth]{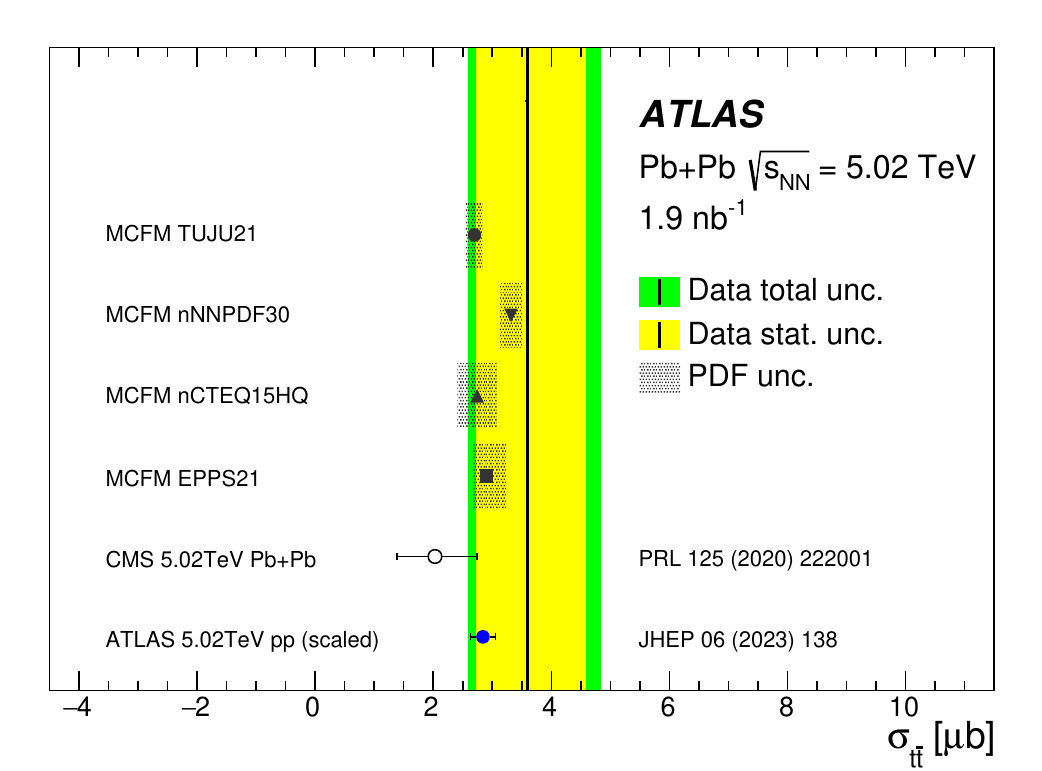}
\caption{Comparison of the $t\bar{t}$ cross section with other measurements and theoretical models based on four nPDF sets~\cite{bib:ttbarPbPb}.}
\label{fig:XSPbPb}
\end{figure}

The $\mu_{t\bar{t}}$ value is determined via a profile-likelihood fit~\cite{bib:fit} to the $m_{e\mu}$ distributions in SR$_1$ and SR$_2$. The measured $\mu_{t\bar{t}}$ is converted into the inclusive $t\bar{t}$ production cross section in Pb+Pb collisions at $\sqrt{s_\mathrm{NN}}=5.02$~TeV in the 0--100\% centrality class, yielding $\sigma_{t\bar{t}}^\mathrm{Pb+Pb} = 3.6\;^{+1.0}_{-0.9}\;\mathrm{(stat.)}\;^{+0.8}_{-0.5}\;\mathrm{(syst.)}~\mathrm{\upmu b} = 3.6\;^{+1.2}_{-1.0}\;\mathrm{(tot.)}~\mathrm{\upmu b}$. The total relative uncertainty of 31\% is dominated by the statistical contribution of 26\%. The observed significance of 5.0 standard deviations constitutes the first observation of $t\bar{t}$ production in Pb+Pb collisions. Figure~\ref{fig:XSPbPb} presents a comparison of the measured $t\bar{t}$ cross section with other experimental results and theoretical predictions. The result is consistent with the measurement in Pb+Pb collisions by CMS~\cite{bib:ttbarPbPbCMS} and $pp$ collisions at $\sqrt{s}=5.02$~TeV by ATLAS~\cite{bib:ttbarpp5TeVATLAS}, scaled to the Pb+Pb system by $A_\mathrm{Pb}^2$. The measurement also agrees with calculations using four nPDF sets: TUJU21, nNNPDF3.0, nCTEQ15HQ, and EPPS21.

\FloatBarrier
%-------------------------------------------------------------------------------
\section{Conclusions}
\label{sec:conclusions}
%-------------------------------------------------------------------------------

Top-quark pair production has been studied in $p$+Pb collisions at $\sqrt{s_{\mathrm{NN}}}=8.16$~TeV with the ATLAS detector. The inclusive $t\bar{t}$ production cross section is measured with a total relative uncertainty of 9\%. The nuclear modification factor is also extracted for the first time for the $t\bar{t}$ process. The results are consistent with other measurements and theoretical predictions from four nPDF sets.

The $t\bar{t}$ process is observed for the first time in Pb+Pb collisions at $\sqrt{s_{\mathrm{NN}}}=5.02$~TeV, with a significance of 5.0 standard deviations. The inclusive $t\bar{t}$ production cross section is determined with a total relative uncertainty of 31\%. Good agreement is found with other experimental results and predictions based on four nPDF sets.

The presented results pave the way for future studies of nPDFs and QGP properties at the LHC. The precision of the $t\bar{t}$ production measurement in $p$+Pb collisions offers new input for constraining nPDFs at high Bjorken-$x$. The first observation of the $t\bar{t}$ process in Pb+Pb collisions opens new opportunities to explore the time evolution of the QGP through hadronic $W$-boson decays originating from top quarks.

%-------------------------------------------------------------------------------
\section{Acknowledgements}
\label{sec:acknowledgements}
%-------------------------------------------------------------------------------

This work was partly supported by the National Science Centre of Poland under grants 2020/37/B/ST2/01043 and 2024/53/N/ST2/00869, by program ``Excellence initiative – research university'' project no 9722 for the AGH University of Krakow, and by PL-Grid Infrastructure.

\bibliography{bibliography}

\end{document}